\documentclass[12pt]{article}

\usepackage{amssymb}
\usepackage{amsmath}
\usepackage{amsfonts}
\usepackage{graphicx}
\usepackage{mathrsfs}
\usepackage{comment}
\usepackage{cite}

\begin{document}

\renewcommand{\figurename}{Fig.}
\renewcommand{\tablename}{Table.}
\newcommand{\Slash}[1]{{\ooalign{\hfil#1\hfil\crcr\raise.167ex\hbox{/}}}}
\newcommand{\bra}[1]{ \langle {#1} | }
\newcommand{\ket}[1]{ | {#1} \rangle }
\newcommand{\beq}{\begin{equation}}  \newcommand{\eeq}{\end{equation}}
\newcommand{\bef}{\begin{figure}}  \newcommand{\eef}{\end{figure}}
\newcommand{\bec}{\begin{center}}  \newcommand{\eec}{\end{center}}
\newcommand{\non}{\nonumber}  \newcommand{\eqn}[1]{\begin{equation} {#1}\end{equation}}
\newcommand{\laq}[1]{\label{eq:#1}}  
\newcommand{\dd}[1]{{d \o d{#1}}}
\newcommand{\Eq}[1]{Eq.(\ref{eq:#1})}
\newcommand{\Eqs}[1]{Eqs.(\ref{eq:#1})}
\newcommand{\eq}[1]{(\ref{eq:#1})}
\newcommand{\Sec}[1]{Sec.\ref{chap:#1}}
\newcommand{\ab}[1]{\left|{#1}\right|}
\newcommand{\vev}[1]{ \left\langle {#1} \right\rangle }
\newcommand{\bs}[1]{ {\boldsymbol {#1}} }
\newcommand{\lac}[1]{\label{chap:#1}}
\newcommand{\SU}[1]{{\rm SU{#1} } }
\newcommand{\SO}[1]{{\rm SO{#1}} }
\def\({\left(}
\def\){\right)}
\def\dt{{d \o dt}}
\def\diag{\mathop{\rm diag}\nolimits}
\def\Spin{\mathop{\rm Spin}}
\def\O{\mathcal{O}}
\def\U{\mathop{\rm U}}
\def\Sp{\mathop{\rm Sp}}
\def\SL{\mathop{\rm SL}}
\def\tr{\mathop{\rm tr}}
\def\ebq{\end{equation} \begin{equation}}
\newcommand{\OR}{~{\rm or}~}
\newcommand{\AND}{~{\rm and}~}
\newcommand{\EV}{ {\rm \, eV} }
\newcommand{\KEV}{ {\rm \, keV} }
\newcommand{\MEV}{ {\rm \, MeV} }
\newcommand{\GEV}{ {\rm \, GeV} }
\newcommand{\TEV}{ {\rm \, TeV} }
\def\o{\over}
\def\a{\alpha}
\def\b{\beta}
\def\c{\varepsilon}
\def\d{\delta}
\def\e{\epsilon}
\def\f{\phi}
\def\g{\gamma}
\def\h{\theta}
\def\k{\kappa}
\def\l{\lambda}
\def\m{\mu}
\def\n{\nu}
\def\p{\psi}
\def\q{\partial}
\def\r{\rho}
\def\s{\sigma}
\def\t{\tau}
\def\u{\upsilon}
\def\v{\varphi}
\def\w{\omega}
\def\x{\xi}
\def\y{\eta}
\def\z{\zeta}
\def\D{\Delta}
\def\G{\Gamma}
\def\H{\Theta}
\def\L{\Lambda}
\def\F{\Phi}
\def\P{\Psi}
\def\S{\Sigma}
\def\me{\mathrm e}
\def\ol{\overline}
\def\tl{\tilde}
\def\*{\dagger}

\begin{center}

\vspace{1.5cm}

{\Large\bf Small cosmological constant from a peculiar inflaton potential}  
\vspace{1.5cm}

{\bf Wen Yin}

\vspace{12pt}
\vspace{1.5cm}
{\em 

{Department of Physics, Tohoku University, Sendai, Miyagi 980-8578, Japan } }

\vspace{1.5cm}

\abstract{
We propose a novel scenario to explain the small cosmological constant (CC) by a finely tuned  inflaton potential. 
The tuned shape is stable under radiative corrections, and our setup is technically natural. 
The peculiar potential approximately satisfies the following conditions: the inflation is eternal 
 if CC is positive, and not eternal if CC is negative. 
By introducing a slowly varying CC from a positive value to a negative value, the dominant volume of the Universe after the inflation turns out to have a vanishingly small CC.   
The scenario does not require eternal inflation but the e-folding number is exponentially large and the inflation scale should be low enough. The scenario can have a consistent thermal history, but the present equation of state of the Universe is predicted to differ from the one for the  $\L$CDM model. 
A concrete model with a light scalar field is studied. 
}

\end{center}
\clearpage

\setcounter{page}{1}
\setcounter{footnote}{0}

\setcounter{footnote}{0}
\section{Introduction}

One of the long-standing theoretical problems in particle theory and cosmology is the fine-tuning of the cosmological constant (CC)~\cite{Weinberg:1987dv, Weinberg:1988cp}, which is measured as \cite{Aghanim:2018eyx}
\beq \L_{C}\simeq 2.2\times 10^{-3}\EV.\eeq
 The CC problem should be solved by  IR dynamics because even 
the QCD contribution~$(1\GEV)^4/(16\pi^2)$ to the CC should be canceled by an amount of tuning of $\O(10^{-45})$. 
For this, non-trivial dynamics should happen when the Universe is much colder than the QCD scale.

Aside from the anthropic solution~\cite{Weinberg:1987dv, Weinberg:1988cp} (and also works relevant to it \cite{DeSimone:2008bq,Bloch:2019bvc}), there have been several proposals to relax the tuning dynamically around the present Universe. Since a vanishing CC is a critical point for the empty Universe to inflate and contract, 
it was studied in e.g. Refs. \cite{Abbott:1984qf,Banks:1984tw} and recently in Ref.~\cite{Graham:2019bfu} that a slowly varying scalar field can drive the Universe at around the critical point.
Although, such a scenario typically  predicts an empty Universe which may be inconsistent with the big-bang cosmology, the authors in Ref.~\cite{Graham:2019bfu} showed that the Universe can be reheated by the scalar field in the contracting Universe and discussed that the produced plasma induces a bounce of the Universe to get the standard cosmology. 

The inflation paradigm, on the other hand, is widely accepted as a central part of the modern cosmology~\cite{Starobinsky:1980te,Guth:1980zm,Sato:1980yn,Linde:1981mu,Albrecht:1982wi}. 
The inflation is driven by a real scalar field, who slow-rolls around a pseudo-flat-direction of the potential. The almost constant potential energy induces an exponential expansion of the spatial volume of the Universe. 
The slow-roll inflation predicts a flat, homogeneous, and slight anisotropic Universe, which has been confirmed from the temperature fluctuation in the cosmic-microwave background (CMB) data~\cite{Planck:2018jri}. 
The inflationary period is cold since the Gibbons-Hawking temperature $H_{\rm inf}/(2\pi)$~\cite{Gibbons:1977mu} is Planck scale suppressed compared to the (false) vacuum energy scale of the Universe. 
In this paper, we focus on the inflationary period to relax the CC. 

The quantum diffusion also traps the inflaton in the pseudo-flat-regime in a probabilistic way.  If the typical rate for finishing the inflation in a single Hubble patch is smaller than the expansion rate, 
there are always inflating volumes. 
Thus inflation never ends in the entire Universe. 
This is known as eternal inflation~\cite{Linde:1982ur,Steinhardt:1982kg,Vilenkin:1983xq,Linde:1986fc,Linde:1986fd,Goncharov:1987ir} (see also~\cite{Guth:2000ka,Guth:2007ng,Linde:2015edk}). In the eternally inflating Universe, due to the infinities there could be ambiguities in defining the probabilities, which depend on the choice of measures. With certain measures and with dynamical CCs, the explanation of the CC during the eternal inflation has been discussed \cite{DeSimone:2008bq,Giudice:2021viw}.

In this paper, we propose an alternative possibility that the CC is relaxed during non-eternal inflation with the Hubble parameter $H_{\rm inf}\ll \L_{C}$. 
The basic idea is as follows. \begin{itemize}
\item[{\bf 1}] 
We assume that the inflation potential is around such a critical point
that if the minimum of the potential, i.e. the CC, has a positive  value, eternal inflation would take place. 
If negative, the inflation period  would be finite. 
The inflaton potential shape around such a criticality is finely tuned but we can make it stable under radiative corrections, and thus it is technically natural.
\item[{\bf 2}] We assume that the CC decreases from a large and positive value. 
Since the inflaton potential is around the criticality, 
most volume of the Universe finishes inflation when the CC crosses zero. As a consequence, a small CC is exponentially favored in a probabilistic way. 
\end{itemize}
We can have a standard reheating and big-bang cosmology without introducing a measure problem. 

As a concrete QFT model, we introduce a scalar field to scan the CC by taking account of the quantum diffusion of the inflaton and the scanning field. 
We confirm the validity of the scenario analytically in the main part, and numerically by  solving the Fokker-Planck equation with the termination effect of inflation in the Appendix \ref{app:1}

\section{Eternal, non-eternal and critical eternal inflation}
\lac{CI}
\subsection{Review on eternal and non-eternal inflation }
For illustrative purpose, let us consider the following form of the potential for the inflaton $\f$, 
\beq
\laq{Vtot}
V= V_\f[\f] + V_C,
\eeq
where we separate the CC, $V_C,$ from the dynamical part of potential, 
and thus $V_\f$ is vanishing at the potential minimum. 
In this review part, we take $V_C$ as a constant, but it will depend on time and space later.  
In our notation, $V_\f$ has  a vanishing minimum, $V_\f[\f_{\rm min}]=0$. 

Let us focus on the hilltop of the potential. The Taylor series is given by
\beq
V_\f[\f]= V_0 +V_\f'' \frac{\f^2}{2}+ \O(\f^3).
\eeq
Without loss of generality, we have taken the hilltop as the origin of $\f$;
$V_\f'' (<0)$ is the curvature; $V_0$ is defined so that  the potential minimum is vanishing.
There are various models with successful cosmology to get this kind of potential top: modified quartic hilltop inflation e.g.~\cite{Nakayama:2012dw, Guth:2018hsa, Matsui:2020wfx}, multi-natural inflation~\cite{Czerny:2014wza, Czerny:2014xja,Czerny:2014qqa,Higaki:2014sja, Croon:2014dma,Higaki:2015kta, Higaki:2016ydn}, ALP inflation~\cite{Daido:2017wwb, Daido:2017tbr, Takahashi:2019qmh} or heavy QCD axion inflation~\cite{Takahashi:2021tff} et al. 
As we will see that the inflation scale should be too low for the natural inflation~\cite{Freese:1990rb,Adams:1992bn} and other simple hilltop inflation~\cite{Linde:1981mu,Albrecht:1982wi} to provide a consistent CMB data~\cite{Planck:2018jri} in this scenario. But it becomes consistent if we introduce another inflaton/curvaton~\cite{Enqvist:2001zp,Lyth:2001nq, Moroi:2001ct} for explaining the CMB data.

The potential top is so flat that inflation can take place.
The eternal inflation may or may not take place depending on the potential shape.
The inflationary volume inflates and the scale factor $R$ increases with 
\beq
\laq{incon}
R^3\propto e^{\int{3 dt H_{\rm inf} }}.
\eeq
Here $H_{\rm inf}\approx \sqrt{V/{3M^2_{\rm pl}}}|_{\f \simeq 0}$ is the Hubble expansion rate. 

The inflaton cannot stay exactly on the hilltop due to the quantum diffusion in the De-sitter space-time. 
The inflaton undergoes random walks. 
By neglecting the curvature compared with $H_{\rm inf}$, the random walk is represented by 
\beq
\dot{\vev{\D \f^2}} \sim  \frac{H^3_{\rm inf}}{(2\pi)^2}.
\eeq
where $\dot{X} $ denotes the (cosmic) time derivative of $X$. 
Due to the diffusion effect, the classical inflaton field value has a probabilistic distribution. 
If the inflationary regime, $|\f| <\f_{\rm inf}$, has a sufficiently flat potential, the distribution of $\f$ during inflation approaches to a constant, $\sim 1/\f_{\rm inf}$.  
The inflaton rolls out of the inflationary range at a probability $ \dot{\f}/\f_{\rm inf}\sim V_\f'' /3H_{\rm inf}$. 
Thus we get the probability that $\f$ remains in $|\f| \leq \f_{\rm inf}$, 
\beq
\laq{oucon}
P\propto e^{\int{ dt C\frac{V_\f ''}{3H_{\rm inf}} }}
\eeq
Here the model-dependent parameter $C =\O(1)$. 
As a result, we find that the inflating volume, $L_{\rm inf}^3$, satisfies 
\beq
\laq{PR}
L^3_{\rm inf}\propto P \cdot R^3\propto e^{\int{(3H_{\rm inf}+C\frac{V_\f ''}{3H_{\rm inf}} )dt}}.
\eeq
If the exponent increases with time, 
\beq
\laq{inf}
3H_{\rm inf} >  C \frac{|V_\f''| }{3H_{\rm inf}}  ~~~~[\rm eternal ~inflation]
\eeq
the inflating volume increases eternally. In other words the second slow-roll condition, $\eta\lesssim 1$, is satisfied at  the top.
This is the well-known condition for eternal inflation. 
It is also known that it may be difficult to discuss the probabilities during the eternal inflation due to the infinities.\footnote{Strictly speaking, the volume distribution, $1/\f_{\rm inf}$ in this regime may not be correct. This is because we cannot neglect the difference of Hubble rate within $|\f|\lesssim \f_{\rm inf}$ for very long inflation. }

If the exponent decreases in time, on the other hand, the total volume decreases. 
\beq
\laq{noninf}
3H_{\rm inf} < C\frac{|V_\f''| }{3H_{\rm inf}}  ~~~~[\text{non-eternal inflation}]
\eeq
Although we need to tune the conditions for inflation to take place at some Hubble patches, the entire Universe finishes the inflation within a finite e-folding. 
The time scale for the termination is $\sim 1/(-3H_{\rm inf}+ C |V_\f''|/(3H_{\rm inf}) ).$
We note that even if the inflation is non-eternal at the hilltop, 
the CMB data can be explained due to a finite period of inflation such as in the inflectionally-point-inflation~\cite{Takahashi:2019qmh, Kadota:2019dol}.

\subsection{Critical eternal inflation and fine-tunings}

We expect that there is always a critical point between 
the eternal and non-eternal regimes by  decreasing $V_C+V_0$, and thus $H_{\rm inf}$, with fixed $V_\f''$:\footnote{To determine the precise value of $C$ we need the higher-order terms of $\f$ as well as the 
detailed study on the quantum diffusion during inflation with $H_{\rm inf}^2\sim |V_\f''|.$  
As long as there is a fixed value of $C$ for the criticality, our conclusion does not change. The determination of $C$ in a specific model will be studied elsewhere.}
\beq
3H_{\rm inf} =C\frac{|V_\f''| }{3H_{\rm inf}}  ~~~~[\text{Critical eternal inflation}].
\eeq
At  the criticality,  the inflation does not end, but the total volume of the inflating Universe does not change. 
As a consequence, the volume of the Universe after inflation approaches infinity but the inflating volume is kept finite. 
We call this kind of inflation, {\it critical eternal inflation}.

In the following, we focus on the possibility that $V_\f$ (but not $V$) drives the inflation very close to this criticality. 
Namely, $V_\f$ should almost satisfy the condition for the critical eternal inflation at $V_C\to 0$, 
\beq
\laq{critical}
\left. 3H_{\rm inf}\right|_{V=V_\f^c} = C \left. \frac{|V_\f''| }{3H_{\rm inf}} \right|_{V=V_\f^c} [\text{Inflaton potential at the criticality}]
 \eeq
Here and hereafter,  $X^c$ denotes the parameter or quantity at the criticality. 

The near-criticality condition is realized by tuning the potential shape while symmetry can stabilize the tuned condition from radiative corrections. For instance, the inflaton  potential can preserve an exact discrete shift symmetry \beq \f\to \f +2\pi f_\f,\eeq 
which may imply that the  inflaton is a pseudo Nambu-Goldstone boson,
 with $f_\f$ being the decay constant. 
 The  potential has a generic form from non-perturbative effects given by 
 \beq
V_\f = \sum_{n=0} \L_n^4 \cos\(\frac{n \f}{f_\f}+\theta_n\) 
 \eeq
with $n$ being the integer and $\theta_n$ is a relativistic phase.
$\L_i^4$ is the order parameter that explicitly breaks the continuous shift symmetry to the discrete one. 
$\L_0^4$ is the constant term that makes the potential vanishing at the potential minimum according to our notation \eq{Vtot}. 
We can tune  $\L_n^4$ and $\theta_n$ to get the potential around the criticality. For instance, we have only a single cosine term with $n=1$, 
the criticality condition suggests $C /(f^c_\f)^2 = 6/M_{\rm pl}^2$ \AND $(\L_0^c)^4= (\L_1^{c})^4=$ arbitrary. 
In addition, the inflaton can have sizable derivative coupling to the SM particles to successfully reheat the Universe. 

To discuss whether the radiative correction will spoil the criticality we can estimate 
 the 1PI effective potential. The Coleman-Weinberg corrections only involving the derivative couplings to the SM particles do not exist.
 From dimensional regularization\footnote{We use the regularization scheme that the discrete shift symmetry is maintained. Note that in any case the quadratic divergence to the potential is absent due to the symmetry. }  the leading contribution is~\cite{Coleman:1973jx},
\begin{align}
V_{CW} \approx\frac{1}{64 \pi^2} V''_\f[\f]^2\left(\ln \frac{ |V''_\f|}{\mu_{\rm RG}^2} - 3/2 \right),
 \end{align}
 where $\m_{\rm RG}$ is the renormalization scale. Since $V_{\rm CW}\sim \O(H_{\rm inf}^4/(8\pi)^2)$ at around the inflationary regime due to the slow-roll condition, 
the near-criticality condition is technically natural up to\footnote{This criticality is expected to exist in, e.g., the  multi-natural inflation models \cite{Czerny:2014wza, Czerny:2014xja,Czerny:2014qqa,Higaki:2014sja, Croon:2014dma,Higaki:2015kta, Higaki:2016ydn, Daido:2017wwb, Daido:2017tbr,  Takahashi:2019qmh,Takahashi:2021tff} which are stable under radiative correction thanks to a discrete shift symmetry. As argued in \cite{Takahashi:2019qmh} 
that the eternal inflation and non-eternal inflation both exists in the  model in two different parameter regimes which are continuously connected.}\beq \laq{technatu} |V_0-V^c_0| \sim\frac{ H_{\rm inf}^4}{(8\pi)^2}.\eeq
As we will see, this quality is enough for our mechanism to work.

Before ending this section, we also comment on the tuning for the initial condition for the inflaton. 
A conservative estimation on the field range for the inflation may be ~\cite{Barenboim:2016mmw}
$
|V_\f' | < H_{\rm inf}^3
$
which is the region where the classical motion is smaller than the quantum diffusion. 
This gives that 
$
\f_{\rm inf} \sim \frac{H_{\rm inf}^3}{V_\f''}.
$
We have to set the inflaton field to be within this range, $|\f| < \f_{\rm inf},$ as the initial condition, which requires a certain amount of tuning. 
If the inflation lasts long enough, this tuning can be compensated by the expanding volume.
As we will see (e.g. Fig. \ref{fig:dist1} in Appendix.\,\ref{app:1}), the initial tuning for inflation as precise as $e^{-10^{\O(10-100)}}$ can be compensated when our mechanism works.\footnote{ In addition, there are several mechanisms to set the correct initial condition~\cite{Linde:1981mu,Albrecht:1982wi, Daido:2017wwb,Takahashi:2019pqf}.}

\section{Small CC from low-scale inflation}

In this section, we take the inflaton potential to be around the criticality $V_\f \simeq V_\f^c.$ 
We show that the CC can be relaxed during inflation if $V_C$ is dynamical. In \Sec{31} we discuss the case 
$V_C$ is time-varying but is constant in space. The discussion provides  several model-independent conditions for our mechanism to work. In \Sec{32}, we  study a specific model for the time-varying $V_C$ by introducing a real scalar field. This leads to a spatially varying $V_C$ by taking account of the diffusion of the scalar field in the De-Sitter space. We present several phenomenological constraints, and the condition to evade the eternal inflation due to the second scalar field. The analytic and intuitive discussion in \Sec{32} is justified by solving the Fokker-Planck equation in Appendix.~\ref{app:1}.

\subsection{Small CC from time-varying potential}
\lac{31}
To present the idea and to provide model-independent conditions for the mechanism, let us first assume that $V_C$ decreases slowly in time but it is a constant in space coordinate, 
i.e.   $V_C=V_C[t]$.  
We take $V_C=V_C^{ i}>0$ and $|\f|<\f_{\rm inf}$ initially. 
We assume for a while that $V_\f=V_\f^c$ for an illustrative purpose. We will come back to relax this assumption. 
For simplicity of analysis, let us focus on the regime $|V_C|\ll V_\f^c[0]$ throughout this paper. 
Then we can expand 
\beq
\laq{Hinf}
H_{\rm inf}\approx  H_{\rm inf}^c+ \frac{V_C[t]}{6M_{\rm pl}^2 H^c_{\rm inf}}.
\eeq
The volume after inflation is produced at a rate
\beq
\laq{tdep}
\dt L^3_{\rm end}\sim  C\frac{|V_\f''|}{3 H_{\rm inf}} L_{\rm inf}^3[t]
\propto 
\exp{ [\int^t{dt' \frac{V_C[t']}{ M_{\rm pl}^2 H^c_{\rm inf}}}]}.
\eeq
Since $V_C$ decreases in time, $V_C$ and time have the one-to-one correspondence. Defining $\k \equiv \dt V_C (<0) $, we obtain 
\beq
\laq{distri}
\frac{d}{d V_{C}} L_{\rm end}^3\propto |\k^{-1}[V_C] | \exp{ [\int_{V_C^{i}}^{V_C}{d x  \frac{\k^{-1}[x] x}{ M_{\rm pl}^2 H^c_{\rm inf}}}]}.
\eeq
This gives the differential distribution of the CC for the Universe finishing the inflation. 
If we can approximate $\k$ to be a constant, we obtain \beq
\boxed{ \frac{d}{d V_{C}} L_{\rm end}^3\propto \exp{[- \frac{ V_C^2}{2|\k| M_{\rm pl}^2 H^{c}_{\rm inf}}]}}.\eeq 
Interestingly, this is a normal distribution the distribution is peaked at $V_C=0$ with a variance of
\beq
\sigma \equiv \sqrt{{M_{\rm pl}^2 H_{\rm inf}^c\ab{\k}}}.
\eeq 
$\sigma$ should be the typical value of $V_C$ in the Universe. 
We get the variance because the e-folds to end the inflation is $\D N_{\rm end} \sim M_{\rm pl}^2 (H_{\rm inf}^c)^2/(V_C)$ during which $V_C$ still changes. 
The variance $\s$ can be also estimated from $V_C \sim |\k \D N_{\rm end}/H_{\rm inf}^c|$. 
The time scale to end the inflation is estimated as 
\beq
\D N_{\rm end} \sim \k^{-1/2}{M_{\rm pl} (H_{\rm inf}^c )^{3/2}} \sim \frac{M_{\rm pl}^2 (H_{\rm inf}^c )^2}{\s }. 
\eeq
The e-folds also represent the time-scale that the change of $V_C$ by $\s$ becomes sensitive to the volume distribution.\footnote{The relation may be understood similarly to the uncertainty principle. If we would like to measure the Hubble expansion rate with an extremely good precision ($\s\to 0$), we need  extremely large e-folds ($\D N_{\rm end}\to \infty$). }

\paragraph{Conditions to relax the CC}

We can relax the CC to the desired value if 
\beq
\s \lesssim \L_C^4.
\eeq
This is the case $V_C$ varies so slow that $\k\lesssim \L_C^8/(M_{\rm pl}^2 H_{\rm inf}^c)$.

So far we have assumed $V_\f=V_\f^c$ for simplicity.
When $V_\f$ is slightly away from $V_\f^c$ by fixing $(V_\f^{c})''=V_\f'',$ 
the deviation $V_0-V_0^c>0 (<0)$ would change the center value of the distribution~\eq{distri} and bias the cosmological constant to a negative (positive) value.
Therefore we need 
\beq
\laq{CCvdef}
|V_0-V_0^c|\lesssim \L_C^4. 
\eeq
 
  In addition, $\f$ diffusion can change the typical energy of the inflaton potential by
 $(V^c_\f)'' (\phi^{c}_{\rm inf})^2.$ 
By requiring this around or smaller than  \beq 
\laq{ultimate}
 \frac{3H_{\rm inf}^4}{(2\pi)^2} \lesssim \L_C^4\eeq
 a model-independent bound on the inflation scale is obtained as $V_0 \lesssim (4\TEV)^4.$
 In other words, the Gibbons-Hawking temperature should be $\lesssim \L_C.$ 
Note that once \Eq{ultimate} is satisfied,  we get the technically natural parameter region with $H_{\rm inf}^4/(8\pi)^2\lesssim |V-V_0^c|\lesssim \L_C^4$ (see also \Eq{technatu}).

In other words, if there is no other contribution, our mechanism naturally predicts the CC of
\beq
\laq{CCtypical}
|\L_C^4|\sim \max{[ \s, \frac{3H_{\rm inf}^4}{(2\pi)^2}, |V_0^c-V_0|]}. 
 \eeq
 with 
$|V_0^c-V_0|\gtrsim \frac{H_{\rm inf}^4}{(8\pi)^2}$ for a technical naturalness.

\lac{31}
\subsection{Small CC from a scalar field}

\lac{32}
To have a slowly time-varying $V_C$, we may introduce a dynamical field, $a$, which slow-rolls during the inflation by $\f$, 
\beq
V_C=V_C[a].
\eeq
In this case, we should take account of $a$ dynamics to check whether it spoils our previous discussion. 
 $a$ is supposed to have a very flat potential due to an approximate continuous shift symmetry. 
We assume again that $V_\f=V_\f^{c}$ for illustrative purpose.

Since the potential is extremely flat, we can expand it around any field value. 
in general, the leading term for $a$  a linear term 
\beq
V_C = V_C' a
\eeq 
where $V_C[0]=0$ is obtained via a field redefinition $a\to a+$constant. 
Let us take $a[0]=a^{i}(>0)$ as the initial condition at $t=0$, and thus $V_C'>0$ for our mechanism to work. 
Then, $a$ undergoes the slow-roll with the classical motion $a^{\rm cl}[t]\approx -t V_C' /3H_{\rm inf}+ a_i$.
$V_C$ rolls down to $\sim 0$ at a time-scale 
\beq
\laq{slowroll}
\D N_{\rm slowroll}[V_C^{i}]\sim  \frac{3(H^c_{\rm inf})^2 V_C^{i}}{(V_C')^2}
\eeq
which will be the longest time scale in this scenario. 
In the terminology of \Sec{31} we obtain 
$
\k\approx -\frac{V_C'^2 }{3H^c_{\rm inf}},
$
$\s \simeq  \frac{|V_C'|M_{\rm pl}}{\sqrt{6}}$ and 
\beq \D N_{\rm end}\simeq  \sqrt{6} \frac{(H^c_{\rm inf})^2 M_{\rm pl}}{|V_C'|}.\eeq

\paragraph{On the quantum diffusion of $a$}

In this model, we have various additional constraints from the quantum diffusion of $a$. (The result relevant to the quantum diffusion is numerically checked by solving the Fokker-Planck equation in Appendix. \ref{app:1}.)
As $\f$,  $a$ undergoes random walk around the trajectory of the classical motion 
\beq
\D a^2[t] \equiv\vev{(a-a_{\rm cl})^2} \simeq t \frac{H^3_{\rm inf}}{(2\pi)^2 }.
\eeq
Here we have assumed that at the beginning all the inflating universe has $a=a_i$. 
Then we get distribution function of $a$ as 
\beq
f[a,t]\propto e^{-\frac{(a-a_{\rm cl}[t])^2}{2\D a^2[t]}}.
\eeq

Notice that this can be obtained when the contribution to  $H_{\rm inf}$ from the quantum diffusion is neglected. 
Since classical motion dominates over the quantum diffusion at the time scale
$
\D N_{\rm diffuse} \sim \frac{9 H_{\rm inf }^6}{(2\pi)^2
 (V_C')^2},
$
we need
 \beq \laq{classicalcond} \D N_{\rm end}\gg \D N_{\rm diffuse }\eeq 
so that the inflation volume is not sensitive to the the quantum diffusion.

With this condition satisfied, we can 
 obtain the volume distribution 
(see Appendix \ref{app:1} for the derivation by Fokker-Planck equation and its more accurate numerical solution)
\beq 
\laq{adis}
\partial_{V_C} L_{\rm inf}^3[t] \propto \partial_a L_{\rm inf}^3[t] \propto P[a,t] \cdot R^3[a,t]\cdot f[a,t] \propto e^{t  \frac{V_C' a}{M_{\rm pl}^2 H_{\rm inf}^c}  -\frac{(a-a_{\rm cl}[t])^2}{2\D a^2}}.
\eeq
At $a=a_i$ we get $  \partial_a L_{\rm inf}^3[t]|_{a=a_i}\propto e^{[\frac{V_C' a_{\rm ini }}{M_{\rm pl}^2 H_{\rm inf}^c}-  \frac{(V_C')^2}{9H_{\rm inf}^5/(2\pi)^2} ]t}$. 
To avoid the inflating volume at $a=a_i$ from dominating over the Universe, and thus to evade the eternal inflation, we obtain 
\beq  
\frac{(V_C')^2}{9(H^c_{\rm inf})^5/(2\pi)^2}  \lesssim \frac{V_{C}^i}{M_{\rm pl}^2 H_{\rm inf}^c}.
\eeq
This gives \beq
\laq{maxrela}
V_C^i \lesssim (10\KEV)^4 \(\frac{1\GEV^4}{V_0}\)^2 \(\frac{V'_C}{10^{-66}\GEV}\)^2.
\eeq
This can be also obtained by requiring that the diffusion $\sqrt{\D a^2}$ is smaller than $\s$ when $a_{\rm cl}$ cross zero: $ V_C' \sqrt{N_{\text{slow-roll}}} \frac{H_{\rm inf}^0}{2\pi}\lesssim \sigma$.  
Thus for the inflation scale, $V_0^{1/4}\sim $ MeV, GeV, 100GeV, we can relax the tuning of the cosmological constant by $V_C^i/\L_C^4\sim 10^{50}, 10^{26}, 10^{10}$ with $V'_C=10^{-66}\GEV, $ which is around the experimental bound as will be explained. 

\paragraph{Parameter region and phenomenology}

The parameter region in $H_{\rm inf}-V_C'$ plane is shown in Fig. \ref{fig:para}. The contours denote $\max(V_C^i/\L_C^4)$, i.e. the maximal amount of the relaxation~\eq{maxrela}. 
In the lower gray region, \eq{classicalcond} is not satisfied and our estimation is invalid. 
In the upper green region, the slow-roll is too fast and  $\s>\L_C^4.$ 
The pink region below the lowest dashed contour cannot have $V_C^{i}> \L^4_C.$ 
\begin{figure}[!t]
\begin{center}  
   \includegraphics[width=105mm]{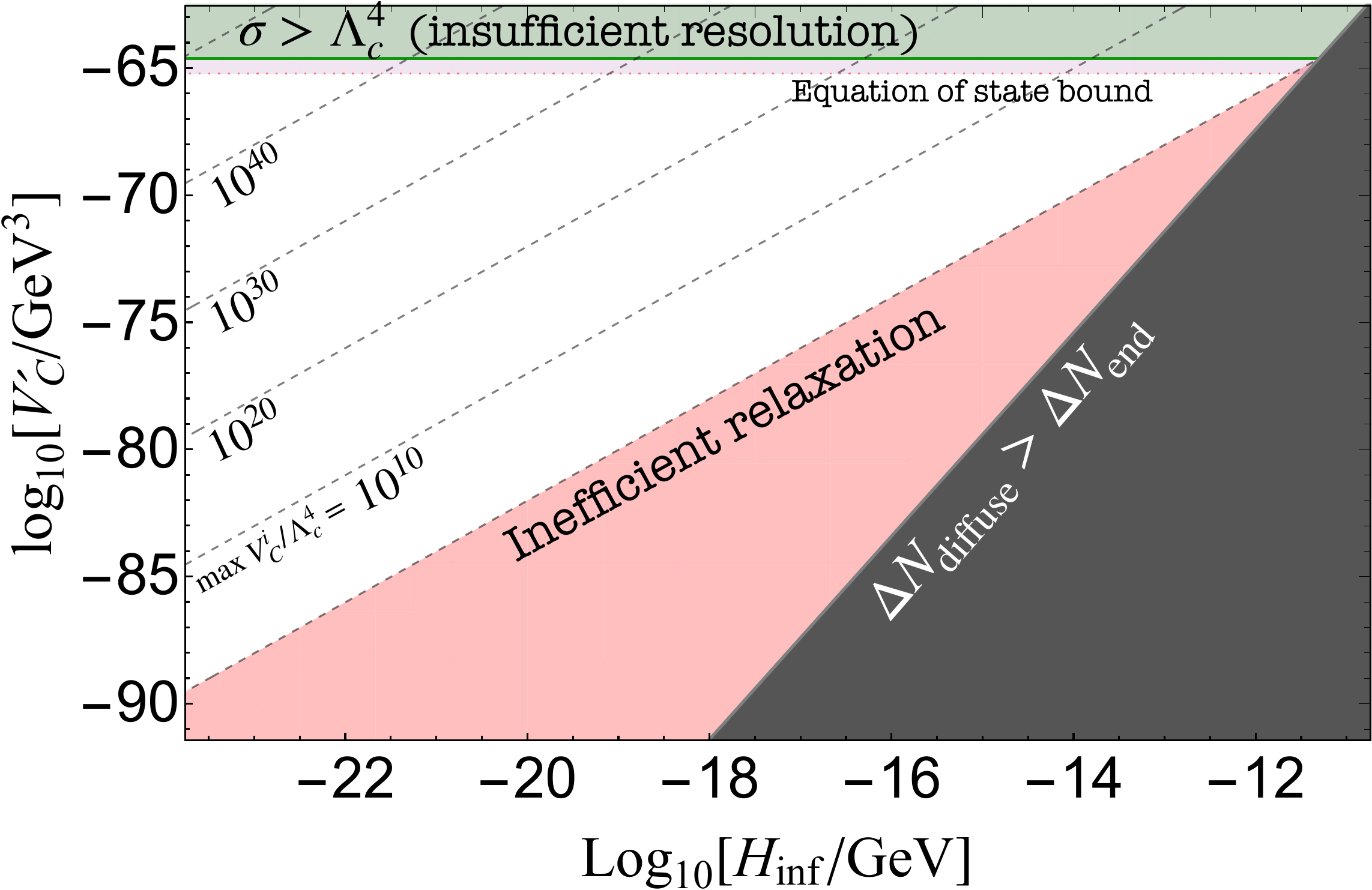}
      \end{center}
\caption{
The contours of $\max V_C^i/\L_C^4$ in $H_{\rm inf}-V_C'$ plane. This is the most efficient relaxation without eternal inflation.
In the lower gray region, our estimation is invalid. In the upper green region, the slow-roll is too fast and the resolution of the relaxation mechanism is worse than $\L_C^4.$ 
The purple region above the dotted line represents the current bound for the slow-rolling $a$. The pink region below the lowest dashed contour cannot have $V_C^{i}> \L^4_C.$
}\label{fig:para} 
\end{figure}

A prediction of this scenario is that the CC is time-varying, 
which leads to the equation of state
$
w\approx \dot{a}^2 \rho_c^{-1}- 1\approx \frac{V_C'^2}{9H^2_{0}} \rho_c^{-1}- 1,
$
with $\rho_c$ being the critical density of the Universe. 
To be consistent with the equation of state~\cite{Planck:2018vyg}, $w < -0.95 (95\%{\rm CL})$,  
we obtain 
\beq
\laq{slowroll}
|V_C'|\lesssim 0.6\times 10^{-65}\GEV^3.
\eeq
Interestingly, this bound is close to the theoretical bound $\s \lesssim \L_C^4 \rightarrow |V'_C| \lesssim 2\times 10^{-65}\GEV^3$ (see appendix.~\ref{app:secinf} for a discussion that $\s \sim \L_C^4$ is favored). 
In particular, the experimental precision will be improved in the Euclid CMB mission~\cite{Amendola:2016saw}, Rubin observatory~\cite{LSSTDarkEnergyScience:2012kar} and DESI~\cite{DESI:2016fyo}, which may probe the scenario. The change of the equation of state linked to the variance of the CC distribution is also the  prediction of the mechanism in~\Sec{31} by assuming a constant $\k$.

The inflationary Hubble parameter should satisfy 
\beq
H_{\rm inf}^c> 2\times 10^{-23}\GEV
\eeq
so that the reheating temperature of the Universe, assuming instantaneous reheating, is larger than $10\MEV$ for successful big-bang nucleosynthesis.

If the inflation scale is low enough, we can obtain $V_C^i$ as large as 
$V_C^i \sim10^{40-50} \L_C^4 $. 
In this case a MeV-GeV scale inflation is required~\cite{Takahashi:2019qmh, Marsh:2019bjr}. 
Since the total e-folds $\D N_{\rm slowroll}$ is exponentially large, 
even very light particles due to an approximate shift symmetry 
reach the equilibrium distribution during the inflation with the energy density of $3H_{\rm inf}^4/(2\pi)^2$ ~\cite{Graham:2018jyp, Guth:2018hsa, Ho:2019ayl}. 
Since this is much smaller than $\L_C^4$, they cannot contribute to a sufficient abundance of the dark matter.
On the other hand, axion dark matter can be produced via the mixing with the axionic inflaton especially if the light axions are at the equilibrium distribution~\cite{Takahashi:2019pqf, Nakagawa:2020eeg}, 
or they can be produced from inflaton decay~\cite{Moroi:2020has,Moroi:2020bkq}. 
The baryogenesis is also possible due to the inflaton decay with higher dimensional operators that are baryon number violating while a proton is stabilized by a parity~\cite{Asaka:2019ocw}.\\

\section{Conclusions}

We have shown that if the inflaton potential has a specific form, 
and if the CC is time-varying, the CC can be relaxed during inflation. 
The price to pay was the tuning, which can be made technically natural, to realize the inflation potential which drives eternal inflation when CC is positive and non-eternal when negative. The Universe is filled by a landscape of the CC with a normal distribution peaked around zero. 
The time-varying CC, if it persists until today, leads to a deviation of the equation of state of the Universe, and can be searched for in the future. In particular, if the measured CC is around the variance of the distribution, the equation of state is predicted to differ from $-1$ by $\O(1-10)\%.$ In a time-varying CC model, the CC can be relaxed from $(10^{3}\GEV)^4$, and in a
slow-rolling scalar model, the CC can be relaxed from $(10\MEV)^4$.


\section*{Acknowledgments}
WY was supported by JSPS KAKENHI Grant Number 20H05851.

\appendix
\section{Second inflation by $a$}
\label{app:secinf}
In the main part we have focused on the inflation driven by $\f$ and its termination. 
We found that most of the Universe finishing the $\f$-inflation has the cosmological constant (CC) almost zero by assuming $\f$ has a peculiar potential form and the CC is time-varying. 

To obtain a consistent CC we need a small enough $|V'_C|$. 
On the other hand, the small $|V'_C|$ can drive the second inflation by $a$ 
if the slow-roll condition 
\beq
\varepsilon(a) \simeq \frac{M_{\rm pl}^2}{2} \left(\frac{V'_C}{V}\right)^2\ll 1
\eeq
is satisfied. 
When the first inflation end with $\tl{V}^i_C \gg \L_C^4$ by chance,  
$V\simeq \tl{V}^i_C \gtrsim \L_{C}^4$. 
Since $\s \simeq  \frac{|V_C'|M_{\rm pl}}{\sqrt{6}} \lesssim \L_C^4$ in our scenario, the second inflation takes place.\footnote{The first inflation reheats the Universe but soon $V_C$ dominates the Universe.  The time-scale for the matter or radiation-dominated Universe can be neglected compared with the inflationary time scales.   }
When $V_C$ decreases to $\sim \s$ the second inflation ends, and then the empty Universe starts to contract.  This Universe may be difficult to have a consistent cosmology unless $\tl{V}_C^{i} \lesssim \L_C^4.$

Here let us estimate the volume distribution of the empty Universe. 
The time-scale of the second inflation can be obtained by solving the slow-roll equation, 
\beq
\D N_{\rm slowroll}^{\rm 2nd}[\tl{V}_C^{i}] \simeq  \frac{1}{2}\frac{(\tl V_C^{i})^2}{(V'_C)^2 M^2_{\rm pl}}.
\eeq
Therefore the volume produced at $t$ satisfying $V_C[t]\simeq \tl{V}_C^i$  increases exponentially by \beq
\laq{2nd}
\D\log{L_{\rm inf,2nd}^3}\sim 3 \D N_{\rm slowroll}^{\rm 2nd}[\tl{V}^i_C] \sim \frac{3}{2} \(\frac{\tl V_C^{i}}{V'_C M_{\rm pl}}\)^2\eeq
due to  the second inflation by $a$. 
On the other hand, the volume undergoing the first inflation also exponentially increases by 
\beq \D\log{L^3_{\rm inf, 1st}} \sim   \frac{3}{2}\( \frac{2\tl{V}^i_C}{(V'_C)^2} \frac{ (V_0-V_0^c)}{ M_{\rm pl}^2}  + \(\frac{\tl V_C^{i}}{V'_C M_{\rm pl}}\)^2\).\eeq 
The second term is the same as \Eq{2nd}. If we would like to require the empty Universe subdominant compared to the Universe with a consistent cosmology, 
we need $V_0 - V_0^c\geq 0$. However, this leads to the negative central value of the CC (see the discussion around \Eq{CCvdef}.) 
Therefore to explain the size of the CC we need other contributions. A candidate is $\s\sim \L_C^4$ from \Eq{CCtypical}, which predicts the current equation of state differing from -1 by $\O(1-10)\%$.\footnote{This may be slightly in tension with the current CMB data, but the Hubble parameter itself may have $\O(10\%)$ tension between the early and late measurement~\cite{DiValentino:2020zio}. 
In any case we can make it consistent if $a$ couples to (dark) particles in the Universe. Then the matter effect easily gives a friction for $a$~\cite{Berera:1995ie,Berera:1998gx,Yokoyama:1998ju, Graham:2019bfu, Nakayama:2021avl}. 
Such a light field then can mediate force and can be tested phenomenologically if it couples to the SM matter or its spin~\cite{Moody:1984ba, Pospelov:1997uv} and if it both couples to SM matter's spin and dark matter~\cite{Kim:2021eye}. The friction is not important during inflation since the (dark) matter is absent there. }

We can alternatively have the positive CC if the $V_C^i$ is close to the upper bound of the contours of Fig.\ref{fig:para}. 
In this case, we have checked numerically that the distribution of $a$ gets broaden due to the expansion effect (see Appendix.\ref{app:1}). 
Also many light particles can contribute the positive CC and further relax the CC (see Appendix.\ref{app:2}).


\section{Solutions to Fokker-Planck equation}

\label{app:1}

Let us explain the dynamics of $a$ during the critical eternal inflation more systematically. 
To this end, we assume that $H_{\rm inf}$ does not change over $\D N_{\rm diffuse}$, 
and estimate the distribution for the ultra-light field, $a$, whose mass can be neglected. 
 The evolution of the classical motion of $a$ is described by the Langevin equation,
\beq
\dot{a}=-\frac{1}{3H_{\rm{inf}}}V'_C(a)+f(\vec{ x} , t),
\laq{Langevin}
\eeq
where $V(a)$ is the potential for $a$ and the dot and prime represent the derivative with respect to the cosmic time $t$ and $a$, respectively.
$f(\vec{x},t)$ satisfies 
\begin{eqnarray}
\left\langle f(\vec{x} , t_1)f(\vec{x} , t_2)\right\rangle=\frac{H_{\rm{inf}}^3}{4\pi^2}\delta(t_1-t_2),
\end{eqnarray}
where $\langle \cdots \rangle$ represents the stochastic average that includes the short wave-length modes. 
The corresponding Fokker-Planck equation can be derived as~\cite{Starobinsky:1986fx,Starobinsky:1994bd}
\begin{eqnarray}
\frac{\partial\mathcal{P}(a, t)}{\partial t}=\frac{1}{3H_{\rm{inf}}}\frac{\partial}{\partial a}(V_C'\mathcal{P}(a, t))+\frac{H_{\rm{inf}}^3}{8\pi^2}\frac{\partial^2\mathcal{P}(a, t)}{\partial a^2},
\end{eqnarray}
where $\mathcal{P}(a, t)$ denotes the probability distribution for the coarse-grained field $a$ in a Hubble patch. This equation describes the time-evolution of $\mathcal{P}(a, t)$ within the time scale, $\D t$, satisfying 
$\D N_{\rm end}>\D t H_{\rm inf}$. 
By assuming a constant $V_C'$, we can solve this equation with an initial condition $P(a, 0)= \d{(a-a_i)}$ as 
\beq
\hat{\mathcal{P}}(a, t,a_i)= \frac{1}{N_{t}} \exp{ \(-\frac{(a-a_{\rm cl}[t,a_i])^2}{2 \D a[t]^2 }\)}.
\eeq
Here $a_{\rm cl}\equiv-t \frac{V'_C}{3H_{\rm inf}} +a_i$ and $\D a^2\equiv t H^3_{\rm inf}/(2\pi)^2$, and $N_{t}\equiv \sqrt{2\pi \D a^2} $ being the normalization factor. 
With a general initial distribution, $\mathcal{P}(a, 0)$, the solution can be obtained from 
\beq
\mathcal{P}(a, t)= \int{d a_i \hat{\mathcal{P}}(a, t,a_i) \mathcal{P}(a_i, 0)}.
\eeq


Let us consider $t\ll \D N_{\rm end}/H_{\rm inf}$ where we have to take account of the back reaction from the  inflaton sector. 
Then it is convenient to follow the volume distribution ${\cal L}^3[a,t],$ by summing over whole $a$ at $t$ we obtain the total volume undergoing the inflation.   
This is usually considered for inflaton field~\cite{Nakao:1988yi, Nambu:1988je, Nambu:1989uf, Linde:1993xx}. 
A similar effect was discussed
in Ref.\,\cite{Graham:2018jyp} for estimating the validity for the estimation of the axion abundance from inflationary equilibrium distribution~\cite{Graham:2018jyp,Guth:2018hsa}. 
Here we use the evolution equation of the inflating volume by taking account of both the inflationary expansion and the terminating probability for the inflation, 
\beq 
\frac{\partial \mathcal{L}^3[a, t]}{\partial t} \approx \(3 H_{\rm inf}-C \frac{V_\f''}{3H_{\rm inf}}\)\mathcal{L}^3[a,t]+
\frac{\partial}{\partial a}\(\frac{V_C'}{3H_{\rm{inf}}} \mathcal{L}^3(a, t)+\frac{H_{\rm{inf}}^{3/2}}{8\pi^2}\frac{\partial H_{\rm{inf}}^{3/2} \mathcal{L}^3(a, t)}{\partial a} \)
\eeq
where $\mathcal{L}^3[a,t]=\partial_a L^3_{\rm inf}[ t]$ in \eq{adis}. 
The first term denotes the Hubble expansion minus the rate to end the inflation due to the $\f$-dynamics  discussed in \Sec{CI}.

We can solve the equation numerically. 
In Fig.\ref{fig:dist1} we show the solution of the inflating volume, $\log{{\cal{L}}^3[a,t]}$ with starting from a normal distribution $a=\frac{1}{\sqrt{2\pi \s^2_{\rm ini}}} e^{\frac{-(a-a_i)^2}{2 \s^2_{\rm ini}}} $
 with a variance $\s_{\rm ini}=M_{\rm pl}/\sqrt{8}$. $V_C^i=2\times10^{16} \L_C^4$, $V_C'=10^{-68}\GEV^3 $ and $H_{\rm inf}=10^{-19}\GEV$ are taken for a sample parameter satisfying the conditions we have discussed in the main part. 
 The initial volume is taken to be  a Hubble volume $1/H^3_{\rm inf}.$
\begin{figure}[!t]
\begin{center}  
   \includegraphics[width=105mm]{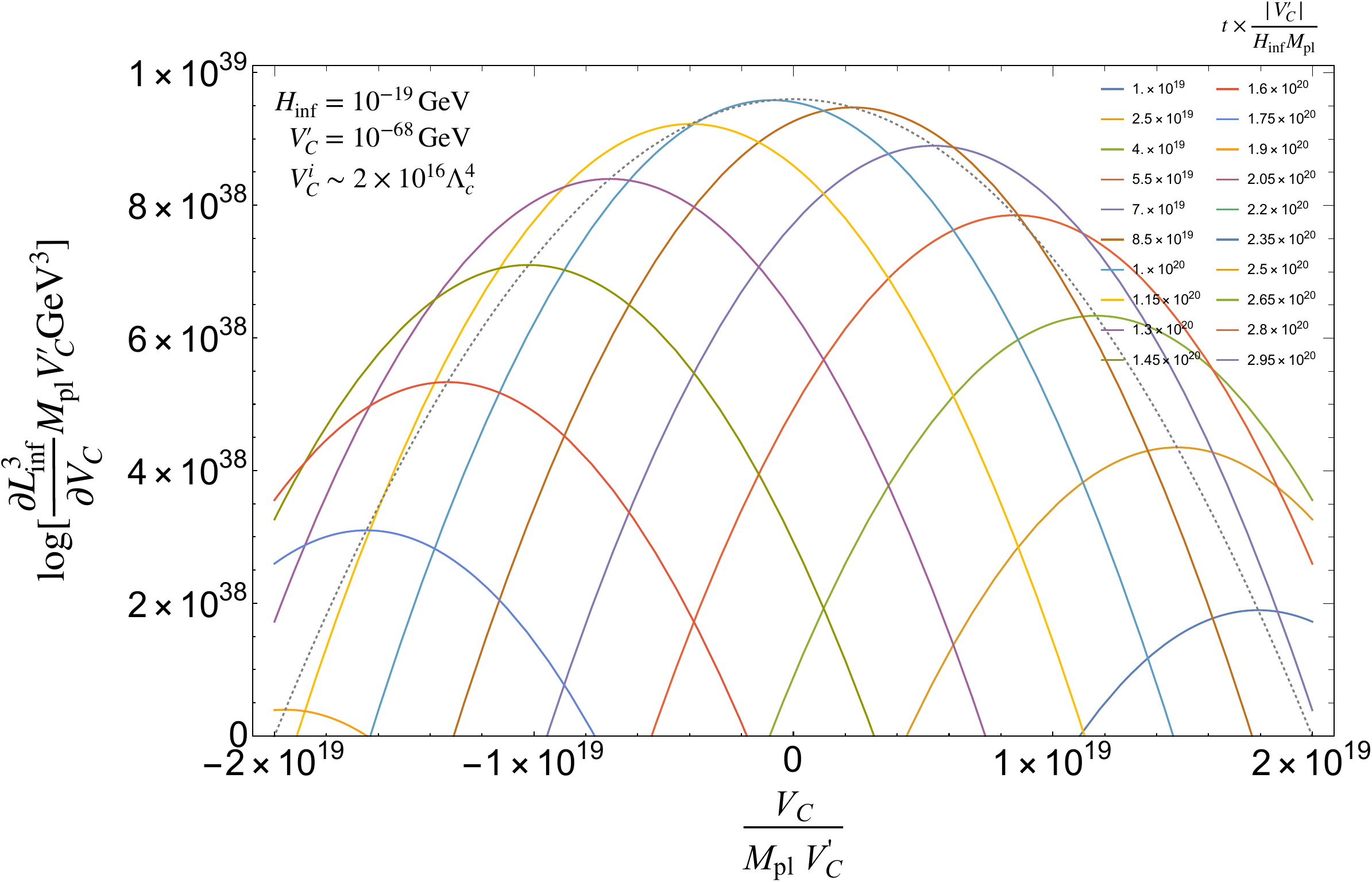}
      \end{center}
\caption{The solution of the Fokker-Planck equation at several $t$ for the solid lines. $V_C^i=2\times10^{16} \L_C^4$, $V_C'=10^{-68}\GEV^3 $ and $H_{\rm inf}=10^{-19}\GEV$ are taken.  The initial distribution is taken as a normal distribution with a variance of $V'_C M_{\rm pl} /\sqrt{8}$. 
The dotted line represents the maximum of the distribution at different $t$. 
 These parameter choices satisfy the conditions we have discussed in the main part. 
 }\label{fig:dist1} 
\end{figure}
We have checked numerically that inflationary volume $\log{{\cal L}^3[a,t]}$ becomes the largest, and thus ${{\cal L}^3[a,t]}$ becomes exponentially largest at $|V_C|< \L_C^4,$ in the whole integration time $t$.

In Fig.~\ref{fig:dist2}, we take $V_C^{i}=10^{26}\L_C^4$ with other parameters unchanged from Fig.~\ref{fig:dist1}. This does not satisfy \eq{maxrela}. 
The inflationary volume is favored at larger $V_C$ as we have explained intuitively and analytically in the main part. Our mechanism does not work in this region. 
\begin{figure}[!t]
\begin{center}  
   \includegraphics[width=105mm]{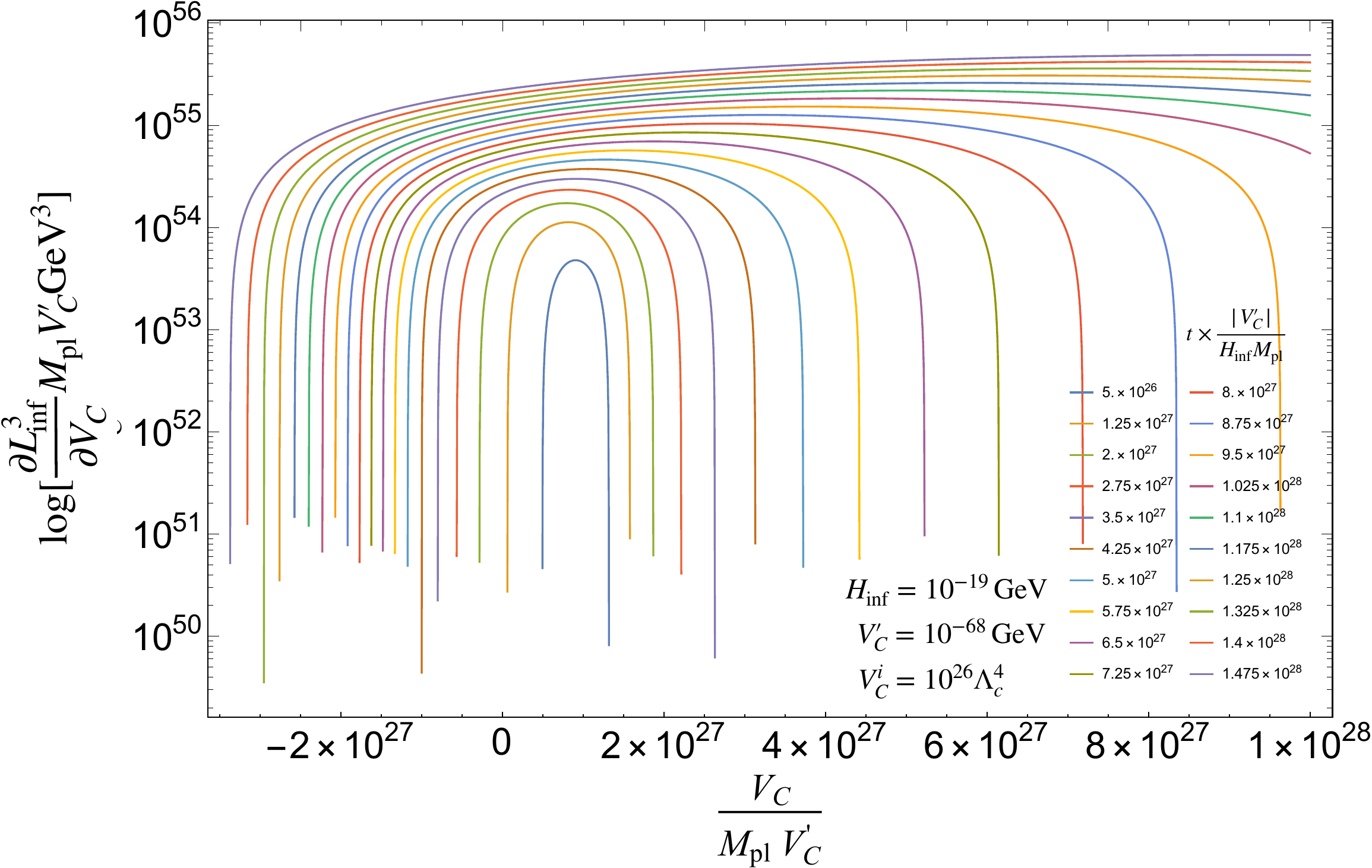}
      \end{center}
\caption{Same as Fig.\,\ref{fig:dist1} except for $V_C^i=10^{26}\L_C^4$, which does not satisfy \eq{maxrela}.
 }\label{fig:dist2} 
\end{figure}

We did not take $\s_{\rm ini}$ much smaller than $M_{\rm pl}$ to have a delta-function due to a numerical limitation. 
Due to the numerical power, we have only checked the boundary of \Eq{maxrela} within a few orders of magnitude. 
Instead we have checked that if $\s_{\rm ini}\lesssim (\gtrsim) \O(1)\s$ with $V_C^i = \O(1) \s$ the mechanism (does not) work.
This is analytically equivalent condition to \Eq{maxrela} (see the discussion below  \Eq{maxrela}). 

\section{Small CC from many scalars} 
\label{app:2}
We can relax the CC by $10^{40-50}$ with  ultra-low-scale inflation with a single scalar. 
We may introduce other particles or interactions to increase the inflation scale or further relax the CC. 

One possible extension of the model 
is to consider $N$ light particles $a_\a$ whose potential is given by $V_{\a}$. 
The vacuum energy  is given by $V_C=\sum_\a{V_{\a}}$. 
We would not obtain a further relaxation of the CC if $V'_\a (>0)$ were a constant. 
This is because the system is equivalent to a single slow-roll field with the linear potential satisfying
\beq
V_{C,eff}'=\sqrt{\sum_{\a=1}^{N}{(V_\a')^2}} 
\eeq
 In other words, all the discussed conditions and constraints apply to the single slow-roll field in the direction of $ -\vec{V}_{\a}'. $
However in a realistic system, a potential $V_\a$ should have a minimum. 
Thus the slow-roll  of $a_\a$ will end when it rolls down and settle into the minimum $a_\a=a_\a^{\rm min}$ at which $V'_{\a}[a_\a^{\rm min}]=0$ but with non-vanishing mass $m_\a^2=V_\a^{''}[a_\a^{\rm min}]>0.$ 

For concreteness, let us assume that $V_{\a}^i- V_\a({a_\a^{\rm min}})=\O(V_{\a}^i)$ 
 the same order $V_{\a}^{i}\sim V_{\b}^{i}=\tl{\L}^4(>0)$ and $V'_{{\a}}\gg V'_{{\a+1}}>0$ at the beginning.  
Then, $a_1$ soon slow rolls to the minimum. When this  happens, the total potential energy $V_C$ decreases by $\O(\tl\L^4)$ from $V_C^{i}=N \O(\tl \L^4).$
After the stabilization of $a_1$, the effective linear potential has $V_{C,eff(1)}'=\sqrt{\sum_{\a=2}^{N}{(V_\a')^2}}\sim V_2'.$ 
This process recursively takes place until $V_C=\O(\tl \L^4)$ which crosses zero during the slow-roll along the direction $\sim a_{n+1}$.
We can neglect the time scales for the stabilization of $a_{\a\leq n}$ compared to the slow-roll timescale of $a_{n+1}$ because of our assumption, $V_\a' \gg V_{\a+1}'$. 

At the last slow-roll, $V'_{C,eff(n)}\sim V_{n+1}', V_{C}\sim \O(\tl\L^4).$ 
Therefore we can use the main part discussion with the replacement of $V'_C\to V_{n+1}'$ and $V_C^{i}\to \O(\tl \L^4)$. 
For $H_{\rm inf}\sim 10^{-18}\GEV$,  $V_{n+1}^{'}=10^{-66}\GEV$ and $n\sim 4\times 10^{24}$, we can relax the CC by $10^{48}$ by assuming $V_C^i=N \O(\tl \L^4)\sim n \O(\tl \L^4)< V_0$. 
We also take into account the equilibrium distribution contribution of $a_{\a\leq n}$ around the minimum. Every $a$ (which is light enough) settled into the minimum forms the equilibrium distribution if $m_\a \ll H_{\rm inf}.$
Thus the energy density is probabilistic with a typical value around $n\times 3H_{\rm inf}^4/(2\pi)^2.$  
We need $n\times 3H_{\rm inf}^4/(2\pi)^2 \lesssim \s$ to prevent the fluctuation of the vacuum energy contributing the volume distribution.


\begin{thebibliography}{99}

\bibitem{Weinberg:1987dv}
S.~Weinberg,
Phys. Rev. Lett. \textbf{59}, 2607 (1987)
doi:10.1103/PhysRevLett.59.2607

\bibitem{Weinberg:1988cp}
S.~Weinberg,
Rev. Mod. Phys. \textbf{61}, 1-23 (1989)
doi:10.1103/RevModPhys.61.1

\bibitem{Aghanim:2018eyx}
N.~Aghanim \textit{et al.} [Planck],
Astron. Astrophys. \textbf{641}, A6 (2020)
doi:10.1051/0004-6361/201833910
[arXiv:1807.06209 [astro-ph.CO]].



\bibitem{DeSimone:2008bq}
A.~De Simone, A.~H.~Guth, M.~P.~Salem and A.~Vilenkin,
Phys. Rev. D \textbf{78}, 063520 (2008)
doi:10.1103/PhysRevD.78.063520
[arXiv:0805.2173 [hep-th]].

\bibitem{Bloch:2019bvc}
I.~M.~Bloch, C.~Cs\'aki, M.~Geller and T.~Volansky,
JHEP \textbf{12}, 191 (2020)
doi:10.1007/JHEP12(2020)191
[arXiv:1912.08840 [hep-ph]].

\bibitem{Abbott:1984qf}
L.~F.~Abbott,
Phys. Lett. B \textbf{150}, 427-430 (1985)
doi:10.1016/0370-2693(85)90459-9

\bibitem{Banks:1984tw}
T.~Banks,
Phys. Rev. Lett. \textbf{52}, 1461-1463 (1984)
doi:10.1103/PhysRevLett.52.1461

\bibitem{Graham:2019bfu}
P.~W.~Graham, D.~E.~Kaplan and S.~Rajendran,
Phys. Rev. D \textbf{100}, no.1, 015048 (2019)
doi:10.1103/PhysRevD.100.015048
[arXiv:1902.06793 [hep-ph]].

\bibitem{Starobinsky:1980te}
A.~A.~Starobinsky,
Phys. Lett. B \textbf{91}, 99-102 (1980)
doi:10.1016/0370-2693(80)90670-X

\bibitem{Guth:1980zm}
A.~H.~Guth,
Phys. Rev. D \textbf{23}, 347-356 (1981)
doi:10.1103/PhysRevD.23.347

\bibitem{Sato:1980yn}
K.~Sato,
Mon. Not. Roy. Astron. Soc. \textbf{195}, 467-479 (1981)
NORDITA-80-29.

\bibitem{Linde:1981mu}
A.~D.~Linde,
Phys. Lett. B \textbf{108}, 389-393 (1982)
doi:10.1016/0370-2693(82)91219-9

\bibitem{Albrecht:1982wi}
A.~Albrecht and P.~J.~Steinhardt,
Phys. Rev. Lett. \textbf{48}, 1220-1223 (1982)
doi:10.1103/PhysRevLett.48.1220

\bibitem{Planck:2018jri}
Y.~Akrami \textit{et al.} [Planck],
Astron. Astrophys. \textbf{641}, A10 (2020)
doi:10.1051/0004-6361/201833887
[arXiv:1807.06211 [astro-ph.CO]].

\bibitem{Gibbons:1977mu}
G.~W.~Gibbons and S.~W.~Hawking,
Phys. Rev. D \textbf{15}, 2738-2751 (1977)
doi:10.1103/PhysRevD.15.2738

\bibitem{Linde:1982ur}
A.~D.~Linde,
Print-82-0554 (CAMBRIDGE).

\bibitem{Steinhardt:1982kg}
P.~J.~Steinhardt,
UPR-0198T.

\bibitem{Vilenkin:1983xq}
A.~Vilenkin,
Phys. Rev. D \textbf{27}, 2848 (1983)
doi:10.1103/PhysRevD.27.2848

\bibitem{Linde:1986fc}
A.~D.~Linde,
Mod. Phys. Lett. A \textbf{1}, 81 (1986)
doi:10.1142/S0217732386000129

\bibitem{Linde:1986fd}
A.~D.~Linde,
Phys. Lett. B \textbf{175}, 395-400 (1986)
doi:10.1016/0370-2693(86)90611-8

\bibitem{Goncharov:1987ir}
A.~S.~Goncharov, A.~D.~Linde and V.~F.~Mukhanov,
Int. J. Mod. Phys. A \textbf{2}, 561-591 (1987)
doi:10.1142/S0217751X87000211

\bibitem{Guth:2000ka}
A.~H.~Guth,
Phys. Rept. \textbf{333}, 555-574 (2000)
doi:10.1016/S0370-1573(00)00037-5
[arXiv:astro-ph/0002156 [astro-ph]].

\bibitem{Guth:2007ng}
A.~H.~Guth,
J. Phys. A \textbf{40}, 6811-6826 (2007)
doi:10.1088/1751-8113/40/25/S25
[arXiv:hep-th/0702178 [hep-th]].

\bibitem{Linde:2015edk}
A.~Linde,
Rept. Prog. Phys. \textbf{80}, no.2, 022001 (2017)
doi:10.1088/1361-6633/aa50e4
[arXiv:1512.01203 [hep-th]].

\bibitem{Giudice:2021viw}
G.~F.~Giudice, M.~McCullough and T.~You,
[arXiv:2105.08617 [hep-ph]].

\bibitem{Nakayama:2012dw}
K.~Nakayama and F.~Takahashi,
JCAP \textbf{05}, 035 (2012)
doi:10.1088/1475-7516/2012/05/035
[arXiv:1203.0323 [hep-ph]].

\bibitem{Guth:2018hsa}
F.~Takahashi, W.~Yin and A.~H.~Guth,
Phys. Rev. D \textbf{98}, no.1, 015042 (2018)
doi:10.1103/PhysRevD.98.015042
[arXiv:1805.08763 [hep-ph]].

\bibitem{Matsui:2020wfx}
H.~Matsui, F.~Takahashi and W.~Yin,
JHEP \textbf{05}, 154 (2020)
doi:10.1007/JHEP05(2020)154
[arXiv:2001.04464 [hep-ph]].

\bibitem{Czerny:2014wza}
M.~Czerny and F.~Takahashi,
Phys. Lett. B \textbf{733}, 241-246 (2014)
doi:10.1016/j.physletb.2014.04.039
[arXiv:1401.5212 [hep-ph]].

\bibitem{Czerny:2014xja}
M.~Czerny, T.~Higaki and F.~Takahashi,
JHEP \textbf{05}, 144 (2014)
doi:10.1007/JHEP05(2014)144
[arXiv:1403.0410 [hep-ph]].

\bibitem{Czerny:2014qqa}
M.~Czerny, T.~Higaki and F.~Takahashi,
Phys. Lett. B \textbf{734}, 167-172 (2014)
doi:10.1016/j.physletb.2014.05.041
[arXiv:1403.5883 [hep-ph]].

\bibitem{Higaki:2014sja}
T.~Higaki, T.~Kobayashi, O.~Seto and Y.~Yamaguchi,
JCAP \textbf{10}, 025 (2014)
doi:10.1088/1475-7516/2014/10/025
[arXiv:1405.0775 [hep-ph]].

\bibitem{Croon:2014dma}
D.~Croon and V.~Sanz,
JCAP \textbf{02}, 008 (2015)
doi:10.1088/1475-7516/2015/02/008
[arXiv:1411.7809 [hep-ph]].

\bibitem{Higaki:2015kta}
T.~Higaki and F.~Takahashi,
JHEP \textbf{03}, 129 (2015)
doi:10.1007/JHEP03(2015)129
[arXiv:1501.02354 [hep-ph]].

\bibitem{Higaki:2016ydn}
T.~Higaki and Y.~Tatsuta,
JCAP \textbf{07}, 011 (2017)
doi:10.1088/1475-7516/2017/07/011
[arXiv:1611.00808 [hep-th]].

\bibitem{Daido:2017wwb}
R.~Daido, F.~Takahashi and W.~Yin,
JCAP \textbf{05}, 044 (2017)
doi:10.1088/1475-7516/2017/05/044
[arXiv:1702.03284 [hep-ph]].

\bibitem{Daido:2017tbr}
R.~Daido, F.~Takahashi and W.~Yin,
JHEP \textbf{02}, 104 (2018)
doi:10.1007/JHEP02(2018)104
[arXiv:1710.11107 [hep-ph]].

\bibitem{Takahashi:2019qmh}
F.~Takahashi and W.~Yin,
JHEP \textbf{07}, 095 (2019)
doi:10.1007/JHEP07(2019)095
[arXiv:1903.00462 [hep-ph]].

\bibitem{Takahashi:2021tff}
F.~Takahashi and W.~Yin,
[arXiv:2105.10493 [hep-ph]].

\bibitem{Freese:1990rb}
K.~Freese, J.~A.~Frieman and A.~V.~Olinto,
Phys. Rev. Lett. \textbf{65}, 3233-3236 (1990)
doi:10.1103/PhysRevLett.65.3233

\bibitem{Adams:1992bn}
F.~C.~Adams, J.~R.~Bond, K.~Freese, J.~A.~Frieman and A.~V.~Olinto,
Phys. Rev. D \textbf{47}, 426-455 (1993)
doi:10.1103/PhysRevD.47.426
[arXiv:hep-ph/9207245 [hep-ph]].

\bibitem{Enqvist:2001zp}
K.~Enqvist and M.~S.~Sloth,
Nucl. Phys. B \textbf{626}, 395-409 (2002)
doi:10.1016/S0550-3213(02)00043-3
[arXiv:hep-ph/0109214 [hep-ph]].

\bibitem{Lyth:2001nq}
D.~H.~Lyth and D.~Wands,
Phys. Lett. B \textbf{524}, 5-14 (2002)
doi:10.1016/S0370-2693(01)01366-1
[arXiv:hep-ph/0110002 [hep-ph]].

\bibitem{Moroi:2001ct}
T.~Moroi and T.~Takahashi,
Phys. Lett. B \textbf{522}, 215-221 (2001)
[erratum: Phys. Lett. B \textbf{539}, 303-303 (2002)]
doi:10.1016/S0370-2693(01)01295-3
[arXiv:hep-ph/0110096 [hep-ph]].

\bibitem{Kadota:2019dol}
K.~Kadota, C.~S.~Shin, T.~Terada and G.~Tumurtushaa,
JCAP \textbf{01}, 008 (2020)
doi:10.1088/1475-7516/2020/01/008
[arXiv:1910.09460 [hep-th]].

\bibitem{Rudelius:2019cfh}
T.~Rudelius,
JCAP \textbf{08}, 009 (2019)
doi:10.1088/1475-7516/2019/08/009
[arXiv:1905.05198 [hep-th]].

\bibitem{Coleman:1973jx}
S.~R.~Coleman and E.~J.~Weinberg,
Phys. Rev. D \textbf{7}, 1888-1910 (1973)
doi:10.1103/PhysRevD.7.1888

\bibitem{Barenboim:2016mmw}
G.~Barenboim, W.~I.~Park and W.~H.~Kinney,
JCAP \textbf{05}, 030 (2016)
doi:10.1088/1475-7516/2016/05/030
[arXiv:1601.08140 [astro-ph.CO]].

\bibitem{Takahashi:2019pqf}
F.~Takahashi and W.~Yin,
JHEP \textbf{10}, 120 (2019)
doi:10.1007/JHEP10(2019)120
[arXiv:1908.06071 [hep-ph]].

\bibitem{Planck:2018vyg}
N.~Aghanim \textit{et al.} [Planck],
Astron. Astrophys. \textbf{641}, A6 (2020)
doi:10.1051/0004-6361/201833910
[arXiv:1807.06209 [astro-ph.CO]].

\bibitem{Amendola:2016saw}
L.~Amendola, S.~Appleby, A.~Avgoustidis, D.~Bacon, T.~Baker, M.~Baldi, N.~Bartolo, A.~Blanchard, C.~Bonvin and S.~Borgani, \textit{et al.}
Living Rev. Rel. \textbf{21}, no.1, 2 (2018)
doi:10.1007/s41114-017-0010-3
[arXiv:1606.00180 [astro-ph.CO]].

\bibitem{LSSTDarkEnergyScience:2012kar}
A.~Abate \textit{et al.} [LSST Dark Energy Science],
[arXiv:1211.0310 [astro-ph.CO]].

\bibitem{DESI:2016fyo}
A.~Aghamousa \textit{et al.} [DESI],
[arXiv:1611.00036 [astro-ph.IM]].

\bibitem{Marsh:2019bjr}
D.~J.~E.~Marsh and W.~Yin,
JHEP \textbf{01}, 169 (2021)
doi:10.1007/JHEP01(2021)169
[arXiv:1912.08188 [hep-ph]].

\bibitem{Graham:2018jyp}
P.~W.~Graham and A.~Scherlis,
Phys. Rev. D \textbf{98}, no.3, 035017 (2018)
doi:10.1103/PhysRevD.98.035017
[arXiv:1805.07362 [hep-ph]].

\bibitem{Ho:2019ayl}
S.~Y.~Ho, F.~Takahashi and W.~Yin,
JHEP \textbf{04}, 149 (2019)
doi:10.1007/JHEP04(2019)149
[arXiv:1901.01240 [hep-ph]].

\bibitem{Nakagawa:2020eeg}
S.~Nakagawa, F.~Takahashi and W.~Yin,
JCAP \textbf{05}, 004 (2020)
doi:10.1088/1475-7516/2020/05/004
[arXiv:2002.12195 [hep-ph]].

\bibitem{Moroi:2020has}
T.~Moroi and W.~Yin,
JHEP \textbf{03}, 301 (2021)
doi:10.1007/JHEP03(2021)301
[arXiv:2011.09475 [hep-ph]].

\bibitem{Moroi:2020bkq}
T.~Moroi and W.~Yin,
JHEP \textbf{03}, 296 (2021)
doi:10.1007/JHEP03(2021)296
[arXiv:2011.12285 [hep-ph]].

\bibitem{Asaka:2019ocw}
T.~Asaka, H.~Ishida and W.~Yin,
JHEP \textbf{07}, 174 (2020)
doi:10.1007/JHEP07(2020)174
[arXiv:1912.08797 [hep-ph]].

\bibitem{DiValentino:2020zio}
E.~Di Valentino, L.~A.~Anchordoqui, O.~Akarsu, Y.~Ali-Haimoud, L.~Amendola, N.~Arendse, M.~Asgari, M.~Ballardini, S.~Basilakos and E.~Battistelli, \textit{et al.}
Astropart. Phys. \textbf{131} (2021), 102605
doi:10.1016/j.astropartphys.2021.102605
[arXiv:2008.11284 [astro-ph.CO]].
\bibitem{Berera:1995ie}
A.~Berera,
Phys. Rev. Lett. \textbf{75}, 3218-3221 (1995)
doi:10.1103/PhysRevLett.75.3218
[arXiv:astro-ph/9509049 [astro-ph]].

\bibitem{Berera:1998gx}
A.~Berera, M.~Gleiser and R.~O.~Ramos,
Phys. Rev. D \textbf{58}, 123508 (1998)
doi:10.1103/PhysRevD.58.123508
[arXiv:hep-ph/9803394 [hep-ph]].

\bibitem{Yokoyama:1998ju}
J.~Yokoyama and A.~D.~Linde,
Phys. Rev. D \textbf{60}, 083509 (1999)
doi:10.1103/PhysRevD.60.083509
[arXiv:hep-ph/9809409 [hep-ph]].

\bibitem{Nakayama:2021avl}
K.~Nakayama and W.~Yin,
[arXiv:2105.14549 [hep-ph]].

\bibitem{Moody:1984ba}
J.~E.~Moody and F.~Wilczek,
Phys. Rev. D \textbf{30}, 130 (1984)
doi:10.1103/PhysRevD.30.130

\bibitem{Pospelov:1997uv}
M.~Pospelov,
Phys. Rev. D \textbf{58}, 097703 (1998)
doi:10.1103/PhysRevD.58.097703
[arXiv:hep-ph/9707431 [hep-ph]].

\bibitem{Kim:2021eye}
D.~Kim, Y.~Kim, Y.~K.~Semertzidis, Y.~C.~Shin and W.~Yin,
[arXiv:2105.03422 [hep-ph]].

\bibitem{Starobinsky:1986fx}
A.~A.~Starobinsky,
Lect. Notes Phys. \textbf{246}, 107-126 (1986)
doi:10.1007/3-540-16452-9\_6

\bibitem{Starobinsky:1994bd}
A.~A.~Starobinsky and J.~Yokoyama,
Phys. Rev. D \textbf{50}, 6357-6368 (1994)
doi:10.1103/PhysRevD.50.6357
[arXiv:astro-ph/9407016 [astro-ph]].

\bibitem{Nakao:1988yi}
K.~i.~Nakao, Y.~Nambu and M.~Sasaki,
Prog. Theor. Phys. \textbf{80}, 1041 (1988)
doi:10.1143/PTP.80.1041

\bibitem{Nambu:1988je}
Y.~Nambu and M.~Sasaki,
Phys. Lett. B \textbf{219}, 240-246 (1989)
doi:10.1016/0370-2693(89)90385-7

\bibitem{Nambu:1989uf}
Y.~Nambu,
Prog. Theor. Phys. \textbf{81}, 1037 (1989)
doi:10.1143/PTP.81.1037

\bibitem{Linde:1993xx}
A.~D.~Linde, D.~A.~Linde and A.~Mezhlumian,
Phys. Rev. D \textbf{49}, 1783-1826 (1994)
doi:10.1103/PhysRevD.49.1783
[arXiv:gr-qc/9306035 [gr-qc]].
\end{thebibliography}
\end{document}